\documentclass[prl, floatfix, reprint, 
showpacs]{revtex4-1}

\usepackage{graphicx}

\draft 

\begin{document}





\title{Ultrafast polarization conversion with plasmonic crystals} 








\author{M. R. Shcherbakov, P. P. Vabishchevich, V. V. Komarova, T. V. Dolgova, A. A. Fedyanin}
\email{fedyanin@nanolab.phys.msu.ru}
\affiliation{Faculty of Physics, Lomonosov Moscow State University, Moscow 119991, Russia}






\date{\today}

\begin{abstract} Femtosecond-scale polarization state conversion is experimentally found in optical response of a plasmonic nanograting by means of time-resolved polarimetry. Simultaneous measurements of the Stokes parameters as a function of time with an averaging time-gate of 130\,fs reveal a remarkable alteration of polarization state inside a single fs-pulse reflected from a plasmonic crystal. Time-dependent depolarization is experimentally found and described within an analytical model which predicts the four-fold enhancement of the polarization conversion effect with the use of the narrower gate. The effect is attributed to excitation of time-delayed polarization-sensitive surface plasmons with a highly birefringent Fano-type spectral profile.
 \end{abstract}


\maketitle 



Polarization is a property of electromagnetic waves which describes the time-averaged trajectory of the electric field vector at a given point of space. It is a commonplace that the timescale over which the polarization is averaged is usually much greater than the period of a single electromagnetic field oscillation. However, recent works demonstrated that the polarization state could be switched on the sub-picosecond scale by means of elementary excitations in quantum-sized media at subzero temperatures \cite{paul,gansen}. A convenient system for observation of sub-picosecond-lifetime elementary excitations at room temperatures is a modulated surface of a noble metal film where surface plasmon-polaritons (SPPs) are excited. Proved by numerous ultrafast experiments involving femtosecond laser pulse sources \cite{ropers,vengurlekar,vabishchevich} the mean lifetime of a resonantly excited SPP is found to be varying from tens to hundreds of femtoseconds which is mainly defined by radiative losses of the SPPs. On the other hand, remarkable polarization properties of SPP-active media  \cite{bryan, gordon} allows one to use plasmonic nanostructures as polarizers \cite{yu}, waveplates \cite{drezet, shcherbakov} as well as handedness-sensitive chiral media \cite{drezet2} of nanoscale thicknesses. Despite the high polarization performance and distinct temporal characteristics of anisotropic plasmonic materials, the possibility of ultrafast polarization control with plasmon-active media has not been demonstrated yet.

%

In this Letter we use resonantly excited surface plasmons to control the state of polarization (SoP) inside a single sub-picosecond telecom laser pulse reflected from a plasmonic nanograting with enormous spectrally dependent optical anisotropy. Time-dependent non-zero depolarization is found which indicates the sub-130\,fs polarization change inside the pulse. We support the experimental data with an analytic model which predicts the four-fold enhancement of polarization conversion which makes plasmonic crystals a perspective media for ultrafast polarization control.

\begin{figure}[b]
\includegraphics[width=0.5\textwidth]{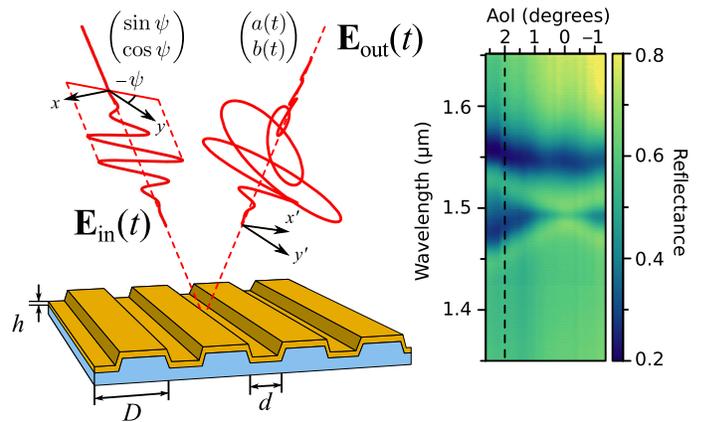}
\caption{\label{sample} (Color online) Left: schematic of the ultrafast polarization conversion with plasmonic crystals. Sample dimensions are $D=1.5\,\mu$m, $d=0.3\,\mu$m, $h=50$\,nm. Right: reflection coefficient of the sample as a function of wavelength and angle of incidence for the $p$-polarized incident light.}
\end{figure}

The concept of ultrafast polarization control with anisotropic plasmonic structures rises from the time-delayed nature of the SPP response and is schematically depicted in Fig.\,\ref{sample}. Incident beam's SoP is split into two linear eigenmodes due to anisotropy of the structure and the $p$-polarized component is time-delayed via exciting SPPs. As a result, sub-pulse polarization alteration is expected in the reflected pulse trains. 

For experimental observation of the ultrafast polarization conversion samples of plasmonic nanogratings were fabricated by thermal sputtering of a gold film onto a polymer diffraction grating made by nanoimprint lithography. The following dimesions of the structure were chosen to match the SPP resonances to the central wavelength $\lambda=1.56\,\mu$m of the femtosecond laser source: period $D=1.5\,\mu$m, grating duty cycle $d/D=0.2$ and film thickness $h=50$\,nm. The reflection spectra of a nanograting sample under different angles of incidence shown in Fig.\,\ref{sample} reveal the minima associated with the band structure of SPPs coupled onto the surface of the film via the 1st and $-1$st diffraction orders. The plasmonic bandgap is seen at the intersection between the plasmonic modes. The bandgap width is $\Delta\lambda\simeq70$\,nm, $\Delta\lambda/\lambda\simeq0.04$ which is comparable to the previous data on spectroscopy of plasmonic crystals \cite{ropers,kitson}. 

\begin{figure}
\includegraphics[width=0.45\textwidth]{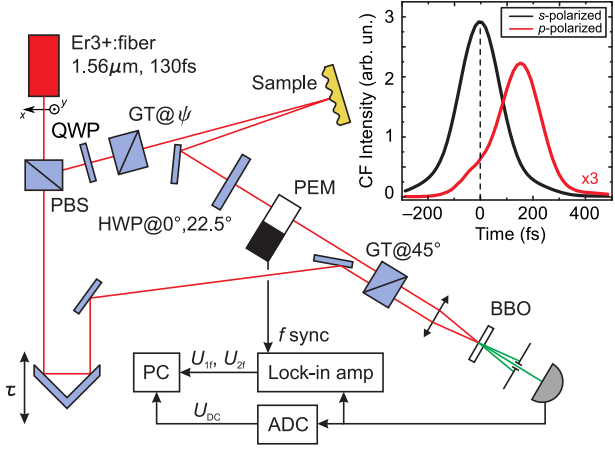}
\caption{\label{stokes} (Color online) Schematic of the setup for time-resolved Stokes parameters measurement. PBS -- polarizing beam splitter, QWP -- quarter-wave plate at 45$^\circ$ with respect to the $Oy$ axis, GT@$\psi$ -- Glan-Taylor polarizer oriented at an angle $\psi$ with respect to the $Oy$ axis, HWP -- half-wave plate, PEM -- photoelastic modulator,  GT@$45^\circ$ -- Glan-Taylor analyzer oriented at an angle $45^\circ$ with respect to the $Oy$ axis. The inset shows the correlation functions measured with the $U_{DC}$ signal at $\psi=0^\circ$ ($s$-polarized, black line) and $\psi=90^\circ$ ($p$-polarized, red line).}
\end{figure}

To begin the ultrafast characterization of the plasmonic crystal time-resolved response of the sample in $s$- and $p$-polarized light was studied to determine the plasmonic impact on the femtosecond pulse shape. An Er$^{3+}$-doped fiber laser (Avesta Project Ltd., Russia) produced a 130\,fs pulse train at a telecom central wavelength of $\lambda=1.56\,\mu$m. The beam was split into two channels one being the signal channel with the sample in it and another being the reference channel with a delay line. Beams of the channels were co-focused onto a BBO crystal under different angles where the non-collinear second harmonic was generated. The non-collinear geometry was chosen to get rid of the collinear second harmonic to observe low intensities at the pulse tails. Correlation functions (CFs) were measured by varying the time delay $\tau$ between the pulses. The results of the CF measurements for $p$- and $s$-polarized incident SoP shown in the inset of Fig.\,\ref{stokes} demonstrate a considerable alteration of the femtosecond pulse shape by the SPPs. A delay of about 180\,fs is seen between the CFs measured for orthogonal input SoPs. 

Symmetry considerations lead horizontally, $Ox$, $p$, and vertically, $Oy$, $s$, polarized states to be the eigenstates of the system under study. If the incident pulse is $s$- or $p$-polarized the SoP of the reflected pulse remains constant as it will be shown below. However, if one sends a linear combination of these states onto the plasmonic crystal the evolution of the SoP inside a single pulse becomes complicated. To experimentally observe the temporal SoP conversion a time-resolved Stokes parameter measurement setup was built. The schematic of the setup is provided in Fig.\,\ref{stokes}. The input SoP is prepared by a quarter-wave plate and a Glan-Taylor prism. The beam reflected from the sample is transformed by a half-wave plate (HWP) and a photoelastic modulator (PEM, Hinds Instruments Inc., USA),  and analyzed by a Glan-Taylor prism. The analyzed pulse contains the temporal distribution of a particular Stokes' vector coordinate at different harmonics of the PEM operating frequency:
\begin{equation}
{\mathbf S}=\left(
\begin{array}{c}
S_0 \\
S_1 \\
S_2 \\
S_3 
\end{array}
\right)=\left(
\begin{array}{c}
U_{DC} \\
U_{2f}(22.5^\circ) \\
U_{2f}(0^\circ) \\
U_{1f}(22.5^\circ) 
\end{array}
\right),
\end{equation}
where the subscripts in the last column indicate the components of the signal modulated at corresponding frequencies and the numbers in brackets denote the angular position of the HWP. $U_{DC}$ is a time-averaged signal measured by an ADC. The pulse containing the information about the Stokes vector components is gated by the reference pulse and detected. CFs measured this way provide the 130\,fs-averaged temporal distribution of the Stokes parameters. This technique allows one to measure the evolution of all the Stokes vector components of the beam reflected from the sample in only two delay line scans.

\begin{figure}
\includegraphics[width=0.45\textwidth]{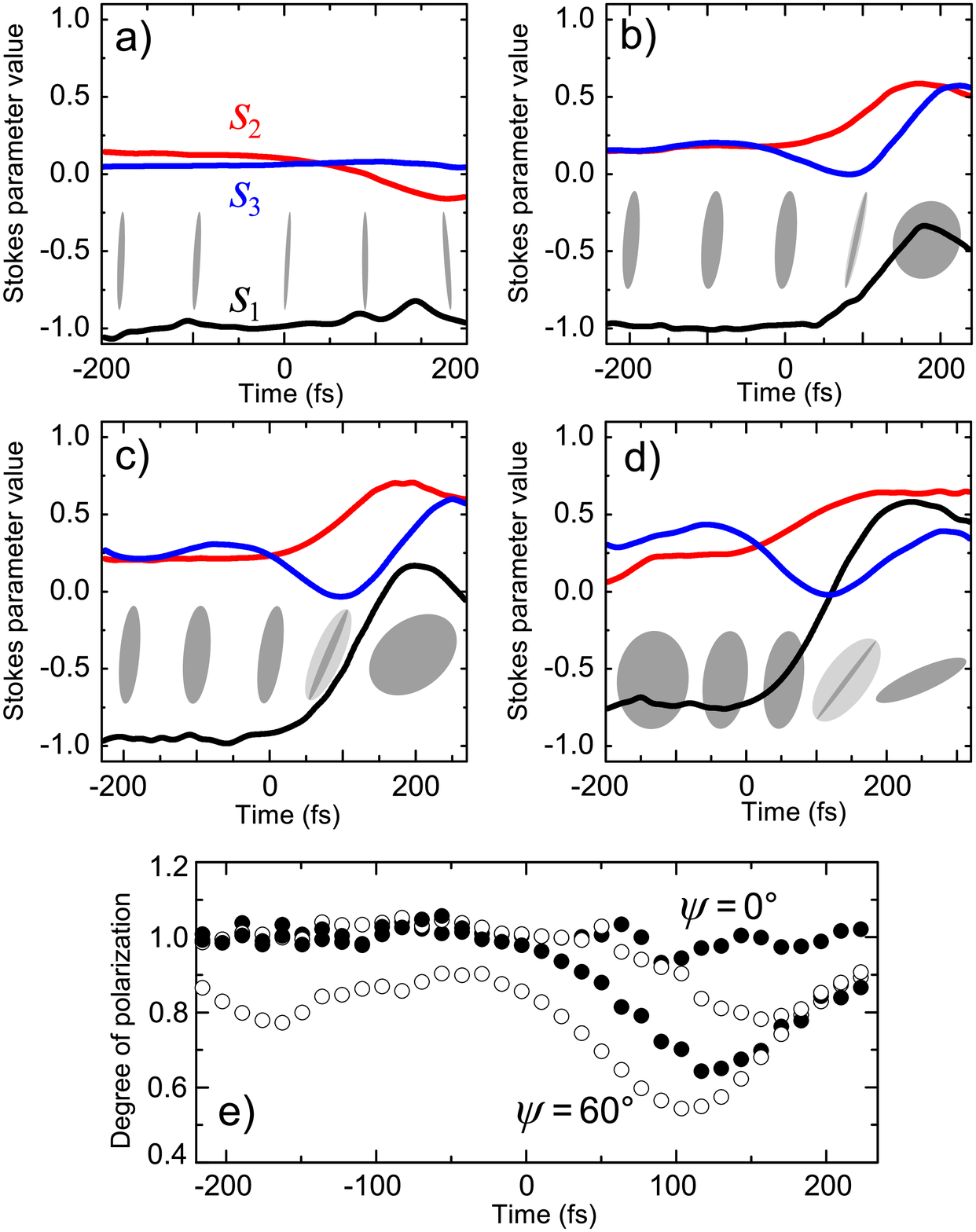}
\caption{\label{tres} (Color online) a)-d) Experimental time-resolved normalized Stokes parameters $s_i$ of the pulse reflected from the nanograting for different incident SoPs with $\psi=0^\circ,30^\circ,45^\circ,60^\circ$, correspondingly. Series of ellipses illustrate the mean trajectory of the E-field vector at moments of time $\tau=-200$\,fs, $-100$\,fs, $0$\,fs, $100$\,fs and $200$\,fs from left to right, correspondingly. Grey stands for the coherent component of SoP and light grey stands for the depolarized one. The sign of $s_3$ defines the handedness of the SoP---right-handed for positive and left-handed for negative values of $s_3$. e) Time-resolved degree of polarization of the reflected pulse SoP for the aforementioned $\psi$ values.}
\end{figure}

Fig.\,\ref{tres} shows the evolution of the polarization of the light pulses reflected from the sample for different input linearly polarized SoPs oriented at an angle $\psi$ with respect to the $Oy$ axis in terms of normalized Stokes parameters $s_i=S_i/S_0$. For $\psi=0^\circ$ the SoP is a linearly $Oy$-polarized state for any moment of time within the uncertainty of the method which is 0.1 on the scale of Stokes parameters. As $\psi$ is increased one can see the modification of the SoP towards the end of the pulse. Various polarization transformations are observed at femtosecond timescale, e.g., at $\psi=30^\circ$ switching between linearly polarized (aspect ratio 0.1) and right-hand circularly polarized (aspect ratio 0.8) is achieved in 100\,fs. Depolarization induced by the sample is also seen in time dependences of degree of polarization defined as $\text{DoP}=\sqrt{S_1^2+S_2^2+S_3^2}/S_0$ in Fig.\,\ref{tres}e.

\begin{figure}
\includegraphics[width=0.47\textwidth]{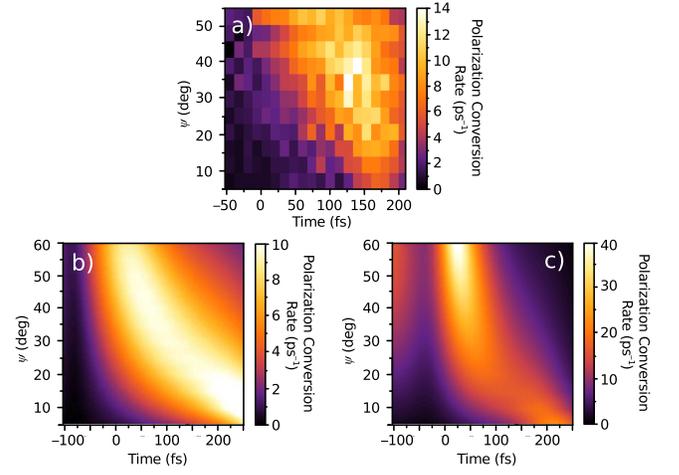}
\caption{\label{rate} (Color online) a) Experimental polarization conversion rate as a function of $\tau$ and $\psi$. The maximum value of the rate equals 13\,ps$^{-1}$ at $\psi=45^\circ$ and $\tau=125$\,fs. b),c) Calculated polarization conversion rate with gates of 130\,fs and 10\,fs, correspondingly.}
\end{figure}

The efficiency of the polarization conversion could be determined by the rate at which the polarization is transformed in the Stokes vector space. It is defined as:
\begin{equation}
v=\left|{\mathbf s}^\prime(t)\right|=
\sqrt{
\sum_{i=1}^{3}
\left(\frac{\partial{s_i}}{\partial{t}}\right)^2
}
\end{equation}
and it represents the speed at which the end of the Stokes vector is moving in the Stokes vector space. Its experimental map is displayed in Fig.\,\ref{rate}a as a function of $\tau$ and $\psi$. The maximum value of the polarization conversion rate is experimentally found to be $v=13$\,ps$^{-1}$ which is twice the maximum rate achieved in 300\,nm-thick Bragg-spaced quantum wells at 80\,K \cite{paul}.

\begin{figure}
\includegraphics[width=0.45\textwidth]{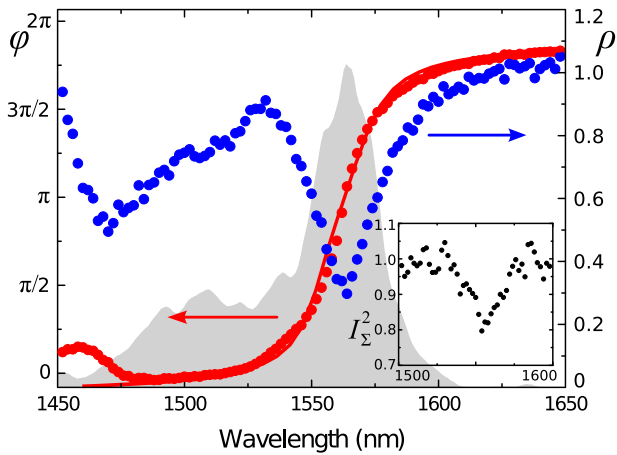}
\caption{\label{ellip} (Color online) Spectra of birefringence $\varphi(\lambda)$ (red dots) and dichroism $\rho(\lambda)$ (blue dots) of the plasmonic crystal at 2$^\circ$-incidence. The grey area shows the spectrum of the pulses used in the ultrafast measurements. The inset shows the sum $I^2_\Sigma$ which is the measure of  depolarization caused by the sample.}
\end{figure}

The effect of the ultrafast polarization conversion and time-dependent depolarization is explained in terms of spectral response of the plasmonic crystal. We characterize optical anisotropy of the sample in CW using the spectrally-resolved ellipsometry setup based on a PEM \cite{jasperson}. The ellipsometric parameters $\rho(\lambda)$ and $\varphi(\lambda)$ of the interface with the following Jones matrix in the basis of horizontally, $(1,0)$, and vertically, $(0,1)$, polarized waves:
\begin{equation}
J(\lambda)\propto\left(
\begin{array}{cc}
 \rho(\lambda)\exp{[\imath \varphi(\lambda)]} & 0 \\
   0 & 1
\end{array}
\right)
\label{jones}
\end{equation}
are measured. Here $\rho(\lambda)$ stands for the ratio between the field reflection coefficients of $p$-polarized and $s$-polarized light and represents linear dichroism; $\varphi(\lambda)$ is the phase delay introduced into one of these states and represents linear birefringence. The formula also applies to an optically anisotropic interface if its optical axis is oriented parallel or perpendicular to the plane of incidence. The spectra $\rho(\lambda)$ and $\varphi(\lambda)$ of the sample are presented in Fig.\,\ref{ellip} for the $2^\circ$-incidence. The dichroism spectrum $\rho(\lambda)$ demonstrates two minima corresponding to the reflection minima at the bandgap edges. The phase delay spectrum $\varphi(\lambda)$ experiences a jump from 0 to 1.8$\pi$ at the long-wavelength edge of the bandgap while the short-wavelength edge provides only the $0.2\pi$ deviation. The spectral phase profile could be understood in terms of the Fano-type response of the nanograting \cite{fano2,lukyanchuk}. The long-wavelength part of the phase curve is fitted with the following expression:
\begin{equation}
\varphi(\omega)=\arg{\left(C+\frac{B e^{i \phi}}{\omega-\omega_0+i \gamma}\right)}-\varphi_0.
\label{eq:fano}
\end{equation}
The best fit to the experimental data with the parameters $C=0.05$, $B=0.94$, $\phi=-2.1$, $\omega_0=1.2\cdot10^{15}$ ($\lambda_0\simeq1570$\,nm), $\gamma=8\cdot10^{12}$  ($\Delta\lambda\simeq10$\,nm) and $\varphi_0=0.05$ is shown in Fig.\,\ref{ellip} with solid line. The estimates of the coupling strength between the resonances \cite{ropers} show that at 2$^\circ$-incidence the resonances could be considered as independent ones.

We analyze the ellipsometric data for the depolarization that the sample induces at the output beam. In order to unambiguously determine the values of $\rho$ and $\varphi$ three quantities were measured independently, namely, $I_1=(\rho^2-1)/(\rho^2+1), I_2=2\rho \sin{\varphi}/(\rho^2+1)$, and $I_3=2\rho \cos{\varphi}/(\rho^2+1)$. In a system described as (\ref{jones}) the condition $I^2_\Sigma=I_1^2+I_2^2+I_3^2=1$ holds; the experimental deviation of $I^2_\Sigma$ from unity indicates the depolarization the system brings into the beam. The said sum is extracted from the ellipsometric data and plotted as a function of the wavelength in the inset of Fig.\,\ref{ellip}. Depolarization of light with the plasmonic crystal is seen at the long-wavelength edge of the bandgap. The origin of depolarization lies in the spread of incident beam tangential wave-vector component $\Delta k$ and it becomes significant when the condition $\Delta k l>1$ holds \cite{altewischer}, where $l$ is the SPP mean free path. In CW measurements the incident beam had a divergence of 0.04 which makes $\Delta k l\simeq1.2$ giving the grounds for the depolarization of light with the sample. On the other hand, the divergence of the laser beam used in ultrafast experiments is much smaller making $\Delta k l \simeq 10^{-2}$. It means that the observed time-dependent depolarization is not explained with spatial inhomogeneity of the incident beam.

We now show that depolarization caused by the sample in the ultrafast experiments is of different nature and can be eliminated by use of a narrower gate. Fig.\,\ref{tres}e shows time dependences of DoP for different incident SoPs. For $\psi=0^\circ$ the pulse is fully polarized within the experimental uncertainty while for other $\psi$ values the polarization is lost and then partially regained. The depolarization appears due to averaging of the detected Stokes parameters within the 130\,fs-gate. Non-zero depolarization means that the polarization conversion happens faster than the gate span. To obtain the actual polarization conversion rate the polarization evolution was calculated with the use of two gate width values of 130\,fs and 10\,fs by utilizing the temporal representation of the Fano-type spectral response. The calculations show the four-fold enhancement of the polarization conversion rate from 10\,ps$^{-1}$ to 40\,ps$^{-1}$ with the narrowing of the gate which is shown in Fig.\,\ref{rate}b,c. Finally, the depolarization vanishes down to values of $1-$DoP$=10^{-4}$ with the use of the 10\,fs-gate for all moments of time which means that experimentally observed time-dependent depolarization is a consequence of the SoP averaging. The ultrafast polarization performance could be further optimized by picking the proper parameters of the Fano resonance which, in turn, are easily tuned by the dimensions of the structure comprising the work to be fulfilled. 


To conclude, ultrafast polarization conversion is found in plasmon-active metallic nanogratings by means of time-resolved Stokes polarimetry. Plasmon-induced birefringence and dichroism with a Fano-type spectral shape cause a pronounced shift of the polarization state inside a single femtosecond telecom laser pulse at the maximum rate of 13\,ps$^{-1}$ in the Stokes vector space. We also deliver a simple analytic model which explains experimentally found depolarization inside the pulses and predicts four-fold polarization conversion rate enhancement. The plasmon-active nanostructure under study is a subwavelength-thickness device operating at room temperatures which is a promising candidate for ultrafast polarization control in novel plasmonic circuits and telecommunication devices.

\end{document}